\documentclass[%
 superscriptaddress,
 preprint,
 showpacs,preprintnumbers,
 amsmath,amssymb,
 aps,
 prb,
]{revtex4-1}

\usepackage{graphicx}
\usepackage{dcolumn}
\usepackage{bm}

\usepackage{upgreek}
\usepackage{amsmath}

\begin{document}


\title{High-efficiency optical frequency mixing in an all-dielectric metasurface enabled by multiple bound states in the continuum}

\author{Tingting Liu}
\email{ttliu@usst.edu.cn}
\affiliation{Institute of Photonic Chips, University of Shanghai for Science and Technology, Shanghai 200093, China}
\affiliation{Centre for Artificial-Intelligence Nanophotonics, School of Optical-Electrical and Computer Engineering, University of Shanghai for Science and Technology, Shanghai 200093, China}

\author{Meibao Qin}
\affiliation{School of Physics and Materials Science, Nanchang University, Nanchang 330031, China}

\author{Feng Wu}
\affiliation{School of Optoelectronic Engineering, Guangdong Polytechnic Normal University, Guangzhou 510665, China}

\author{Shuyuan Xiao}
\email{syxiao@ncu.edu.cn}
\affiliation{Institute for Advanced Study, Nanchang University, Nanchang 330031, China}
\affiliation{Jiangxi Key Laboratory for Microscale Interdisciplinary Study, Nanchang University, Nanchang 330031, China}

\begin{abstract}
	
We present nonlinear optical four-wave mixing in a silicon nanodisk dimer metasurface. Under the oblique incident plane waves, the designed metasurface exhibits a multi-resonant feature with simultaneous excitations of three quasi-bound states in the continuum (BIC). Through employing these quasi-BICs with maximizing electric field energy at the input bump wavelengths, significant enhancement of third-order nonlinear processes including third-harmonic generation, degenerate and non-degenerate four-wave mixing are demonstrated, giving rise to ten new frequencies in the visible wavelengths. This work may lead to a new frontier of ultracompact optical mixer for applications in optical circuitry, ultrasensitive sensing, and quantum nanophotonics.
	
\end{abstract}

\maketitle


\section{\label{sec1}Introduction}

Optical frequency mixing with efficient generation of new optical frequencies has found widespread use in various emerging research areas, including attosecond pulse generation, supercontinuum generation, optical frequency comb generation, material characterization, and quantum optics\cite{Boyd2020}. Amongst the nonlinear frequency conversion processes, four-wave mixing (FWM) is a versatile third-order nonlinear effect where four photons of different frequencies are mixed together with generated frequencies governed by the equation $\omega_{\text{FWM}}=\pm\omega_{1}\pm\omega_{2}\pm\omega_{3}$ in the case of three different incident photon frequencies $\omega_{1}$, $\omega_{2}$, and $\omega_{3}$. Recently, the ability to realize and harness FWM process at subwavelength volume has become a popular pursuit in the community. A substantial progress in nonlinear frequency conversion at the nanoscale is achieved with the use of dielectric resonant metasurfaces\cite{Sain2019, Pertsch2020, Grinblat2021, Liu2022}. The high refractive index of dielectric material entails a large nonlinear susceptibility and the dielectric resonators confine electromagnetic fields inside. The dielectric metasurfaces underpin strong nonlinear response at the nanoscale as well as flexible control of nonlinear scattering, when combined with the interplay between different multipolar resonances such as magnetic resonance\cite{Carletti2015, Gao2018, Frizyuk2019, Liu2021}, Fano resonance\cite{Yang2015, Vabishchevich2018, Ban2018}, and anapole excitation\cite{Grinblat2016, Xu2018}. 

A recently emerged concept of optical bound states in the continuum (BIC) has been developed to enhance light-matter interaction\cite{Hsu2016, Sadreev2021, Huang2023}. It provides an artificially intriguing approach to strongly confine and localize electromagnetic waves into subwavelength scale via multipolar control, which permits a direct link from ultrahigh quality ($Q$) factors excited in BIC dielectric metasurfaces to high-efficient nonlinear frequency conversion on chip. The electromagnetically engineered quasi-BIC supported by dielectric metasurfaces has been successfully utilized to enhance the second- (SHG)\cite{Carletti2018, Koshelev2020, Anthur2020, Ning2021, Huang2021, Zheng2022}, third- (THG)\cite{Xu2019, Koshelev2019, Liu2019, Gandolfi2021, Sinev2021}, and high-harmonic generation (HHG) processes\cite{Carletti2019, Xiao2022, Xiao2022a}, up to a few orders of magnitude. Most recently, the BIC-assisted dielectric metasurface is demonstrated to generate efficiently high optical harmonics up to the 11$^{\text{th}}$ order\cite{Zograf2022}. Different from the harmonic generation processes above, FWM process is involved with different incident wavelengths and thereby the input wavelengths concurrently resonant is required for its efficiency enhancement. Although the resonantly enhanced degenerate FWM have been studied \cite{Grinblat2017, Liu2018, Colom2019, Xu2022}, the non-degenerate FWM involved with three simultaneous resonances is not easily accessible to the metasurface platform. In particular, up until now, enhanced FWM process empowered by the multiple BIC resonances of dielectric metasurfaces has rarely been explored.

In this work, we demonstrate high-efficient FWM processes in a dielectric metasurface by simultaneous excitations of multiple BIC resonances at input wavelengths. In the metasurface composed of silicon nanodisk dimers, the presence of the quasi-BIC resonances excited by $y$-polarized oblique incident plane wave significantly increase the electric energy inside the structure. Together with the three-order nonlinear susceptibility of silicon, frequency mixing processes give rise to ten new frequencies spanning the visible region from $\sim$450 nm to $\sim$550 nm. The peak output power of these processes is achieved when the excitation wavelengths correspond to maximum of the internal electric energies at the quasi-BIC resonances.

\section{\label{sec2}Multiple bound states in the continuum}

Figure 1 shows a schematic of nonlinear optical frequency mixing processes in a silicon metasurface pumped by three laser beams. The metasurface consists of a periodic square array of silicon nanodisk dimers, with a two-fold $C_{2v}$ rotational symmetry. The lattice period is $p=960$ nm. Each nanodisk has radius $r=210$ nm and height $h=280$ nm, and the distance between the two nanodisks is $d=40$ nm. The refractive index of silicon is presented in the Supplemental Material\cite{SM} (see, also, reference \cite{Palik1985} therein). Note that we choose the nanodisk dimer design since it belongs to a symmetric geometry of the $C_{2v}$ group and possesses abundant structural parameters to be adjusted for peculiar electromagnetic excitations. In the previous works, the toroidal dipole BIC\cite{He2018, Zhou2020} and the non-radiative anapole\cite{Rocco2018}, have been found in the similar design. In a more recent work, the nanohole dimer that is complementary to such structure is exploited to support multi-resonance including a quasi-BIC resonance for the degenerate FWM process\cite{Xu2022}.
\begin{figure}[htbp]
	\centering
	\includegraphics[scale=1.50]{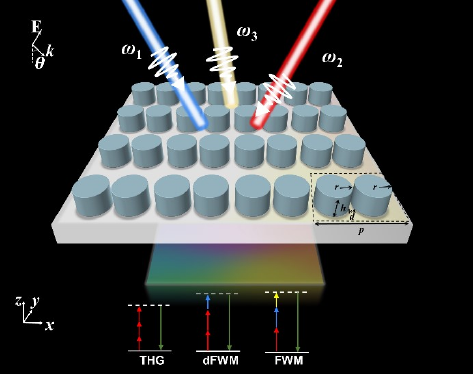}
	\caption{The schematic of the designed metasurface consisting of a square array of silicon nanodisk dimers, simultaneously illuminated by three pump beams termed as $\omega_{1}$, $\omega_{2}$, and $\omega_{3}$ in the near-infrared regime to generate a variety of new visible frequencies. The unit cell is indicated by a dotted square. The bottom inset illustrates the schematic energy diagrams of the nonlinear optical processes that occur simultaneously to generate visible emissions. }
	\label{fig1}
\end{figure}

To boost the nonlinear optical frequency mixing, the designed metasurface simultaneously supports the BIC resonances at the wavelengths of the three input beams in the near-infrared regime. This can be confirmed by performing eigenmode analysis as well as by the simulated electromagnetic field profiles under $y$-polarized normal incidence in the Supplemental Material\cite{SM} (see, also, reference \cite{Hsu2016} therein). Note that the distributions of the electromagnetic fields are even under $C_{2v}$ around the $z$ axis for eigenmodes 2, 3, and odd for mode 1, and thus these modes cannot be simultaneously excited by a normally incident plane wave whose electromagnetic fields are odd. The $y$-polarized inclined plane waves are adopted to introduce the radiation leakage channels to transform the genuine BIC into the quasi-BIC resonances. Fig. 2(a) presents the evolutions of the transmission characteristics of the metasurface when the inclined incident angles are increased. The leaky resonances of quasi-BIC can be visualized with the observable finite linewidth of transmission under the oblique incidence. As a demonstration for resonantly enhanced frequency mixing, an incident angle of 10$^{\circ}$ is adopted here due to a tradeoff between calculation speed and accuracy. In this case, the simulated transmission spectra of the metasurface exhibit the three narrow resonances around the corresponding wavelengths shown in Fig. 2(b). We estimate the $Q$ factors of the quasi-BIC resonances by fitting the simulated transmission spectra with a Fano line shape expressed by $T=|a_{1}+ia_{2}+\frac{b}{\omega-\omega_{0}+i\gamma}|^{2}$, where $a_{1}$, $a_{2}$, and $b$ are real numbers, $\omega_{0}$ is the resonance frequency of quasi-BIC, and $\gamma$ is the total leakage rate\cite{He2018, Li2019, Qin2022}. Meanwhile, the multipolar characteristics of each resonance is quantitatively identified using the Cartesian multipole decomposition, shown in the Supplemental Material\cite{SM} (see, also, reference \cite{Kaelberer2010, Savinov2014} therein). The asymmetric oblique incidence of light triggers the transition from BIC to quasi-BIC reonances via small but perceptible interaction with the continuum, giving rise to the ultrahigh $Q$ factors to increase the field enhancement.
\begin{figure}[htbp]
	\centering
	\includegraphics[scale=0.60]{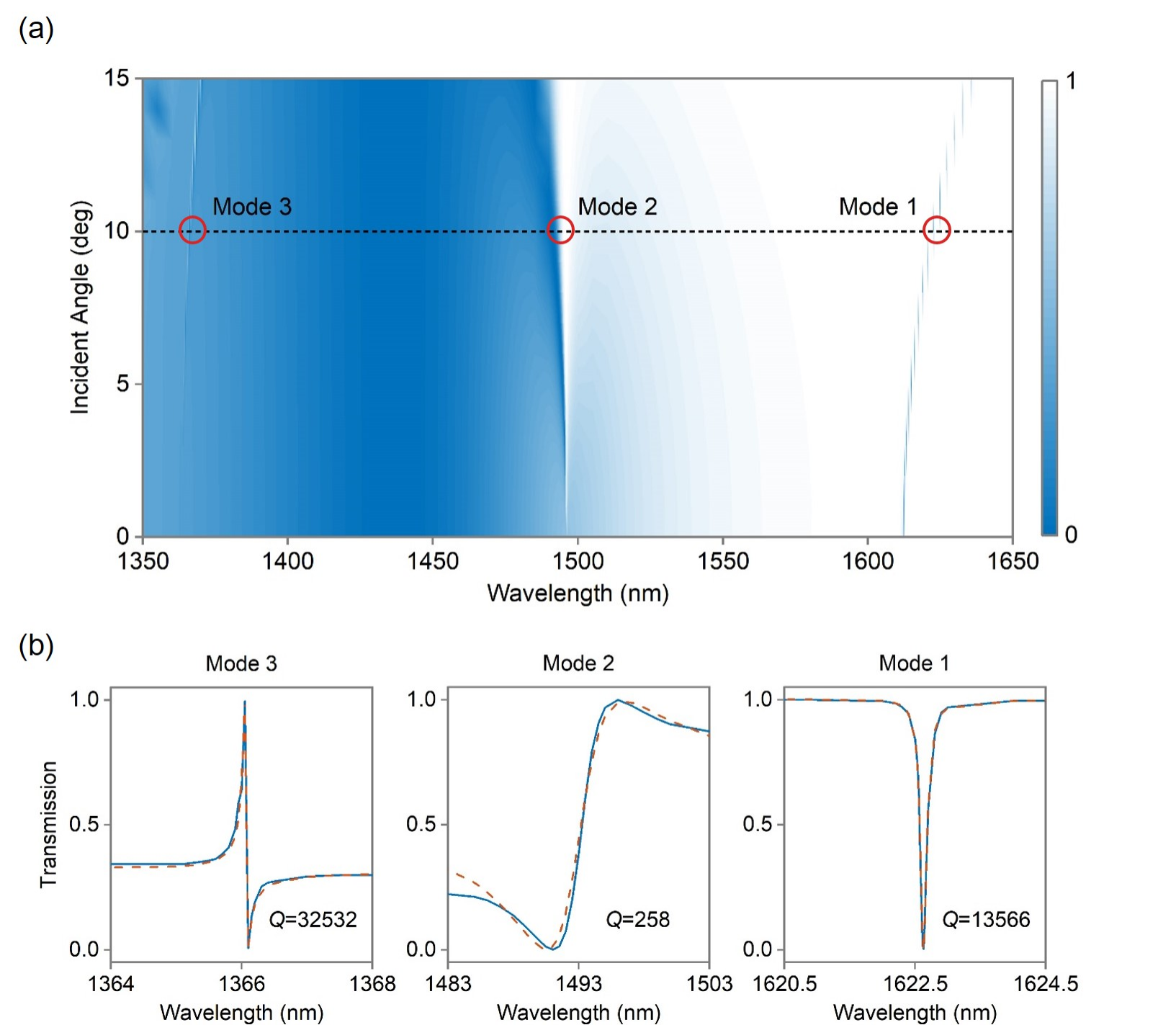}
	\caption{(a) The transmission of the metasurface as a function of wavelength and incident angle. As seen, the resonances corresponding to the excitations of the three quasi-BIC modes show narrow linewidth, when the modes evolve from the perfect BIC to quasi-BIC. (b) The simulated (solid) and theoretical (dashed) transmission spectra of the metasurface under $y$-polarized oblique incidence of 10$^{\circ}$ in the vicinity of the three quasi-BIC resonances.}
	\label{fig2}
\end{figure}

\section{\label{sec3}Nonlinear harmonic generation and frequency mixing}

The nonlinear response is simulated with the frequency domain solver in COMSOL MULTIPHYSICS through two sequential computations. The electric field at the fundamental wavelength are computed in the first step and then a nonlinear polarization density is induced as the source term of the wave equation in the second step. For the silicon material considered here, the nonlinear polarization associated with the third-order nonlinear process can be expressed by\cite{Boyd2020} 
\begin{equation}
	P_{i}(\omega_{1}+\omega_{2}+\omega_{3})=\varepsilon_{0}D\sum_{jkl}\chi_{ijkl}^{(3)}E_{j}(\omega_{1})E_{k}(\omega_{2})E_{l}(\omega_{3}),
	\label{eq1}
\end{equation}
where $\varepsilon_{0}$ is the vacuum permittivity, $D$ is the degeneracy factor, $\chi^{(3)}$ is the third-order nonlinear susceptibility of silicon, and $E$ is the localized electric field at the fundamental wavelength. Here $\chi^{(3)}=2.45\times10^{-19}$ m$^{2}$/V$^{2}$ is adopted for silicon in the near-infrared\cite{Xu2018, Carletti2019}, and the optical intensity of fundamental pump is fixed to 1 MW/cm$^{2}$.

\begin{figure}[htbp]
	\centering
	\includegraphics[scale=0.60]{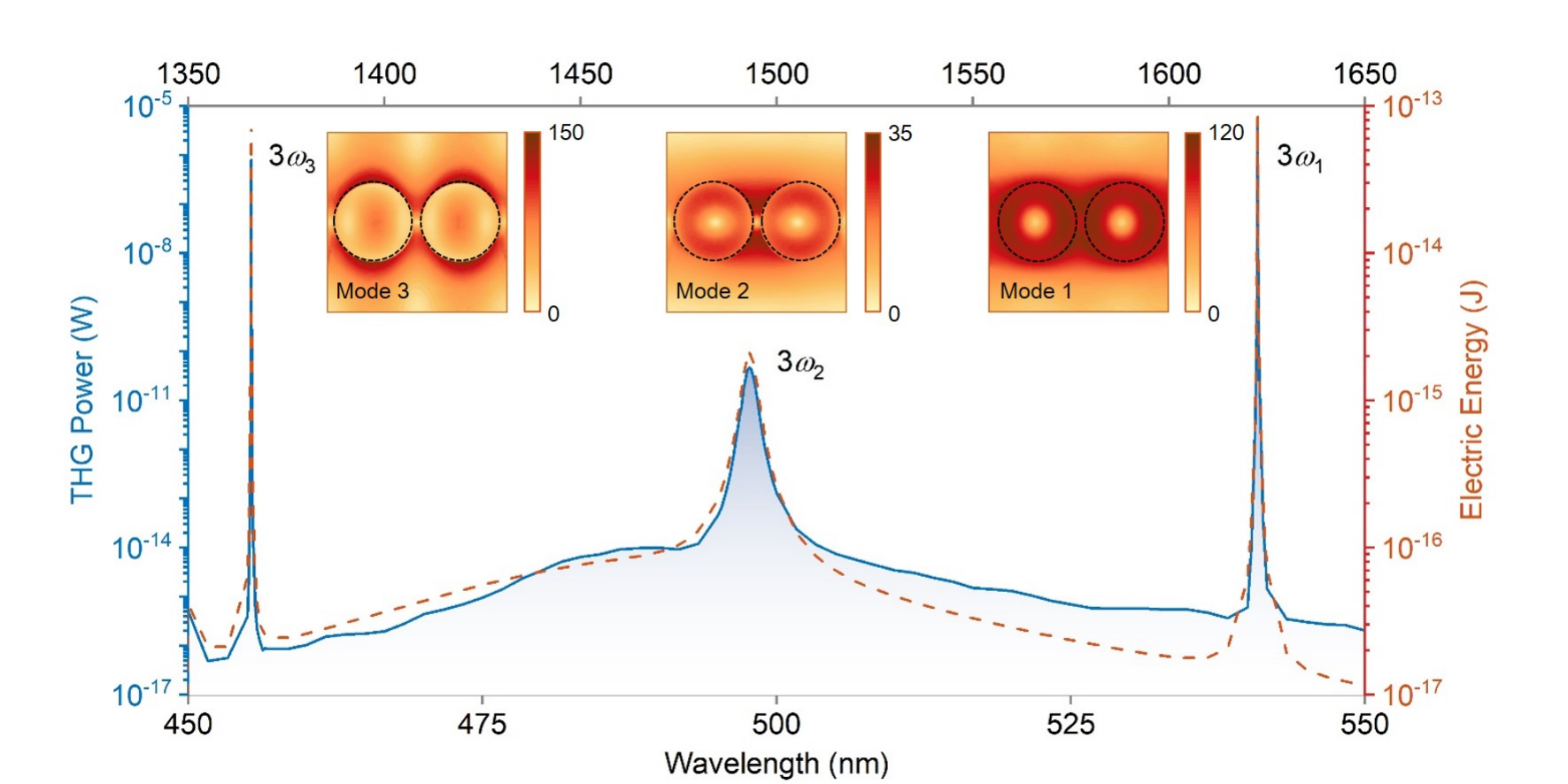}
	\caption{The THG power (solid) and electric energy inside the nanodisk dimers (dashed) of the metasurface under $y$-polarized oblique incidence. Insets are the enhancement of electric field intensity ($|E|/|E_{0}|$) at the fundamental wavelengths of the three quasi-BIC resonances.}
	\label{fig3}
\end{figure}
To start with, we study the harmonic generations by the metasurface when pumped by a single near-infrared beam. In Fig. 3, the envelope of the THG power spectrum presents three distinctive peaks at 455.35 nm, 497.67 nm, and 540.88 nm, respectively. Since the third-order nonlinear effect essentially behaves as a volume phenomenon, we use the volume electric energy at the excitation wavelengths to characterize the excitation power here.  As shown in Fig. 3, the electric energy stored in the resonators reaches the peak values at the quasi-BIC resonance wavelengths corresponding to the modes 1, 2, 3 as anticipated, in consistent with the BIC feature to effectively confine the electric field inside the dielectric resonators. The essential connection between the nonlinear emission and internal electric field energy explains the very good agreement between the three peaks of THG power and the three maxima of electric energy. Intriguingly, although  resonance mode 1 shows lower $Q$-factor and field enhancement at fundamental wavelength 1622.63 nm than those of mode 3 at 1366.05 nm,  the peak of THG power corresponding to mode 1 at the wavelength 540.88 nm is approximately 5 times larger than the peak power of mode 3 at 455.35 nm because of the relatively larger electric energy. Their efficiency difference originates from the fact that the electric field of quasi-BIC resonance at mode 1 is preferentially located inside the nanodisks, rather than that near the edges at mode 3, which is a more beneficial aspect for nonlinear THG processes with large volume contribution.

Due to the presence of multiple quasi-BIC resonances in the designed metasurface, it is possible to study FWM by exciting either single or multiple modes. Justified by the enhanced THG signals in Fig. 3, the combination of highly confined electromagnetic fields within the nanodisk dimers and the strong intrinsic material nonlinearities of silicon should allow FWM process to be effectively exploited. Fig. 4 shows the results of mixing different excitation frequencies at the three quasi-BIC resonances at $\omega_{1}$, $\omega_{2}$, and $\omega_{3}$, respectively. When the three pump beams are temporally coincident, the ten spectral peaks ranging from $\sim$450 nm to $\sim$550 nm can be obtained in the visible window. In Figs. 4(a)-(c) for the degenerate four wave mixing, two pump beams are considered with one at a fixed wavelength of $\omega_{1}$, $\omega_{2}$, and $\omega_{3}$, respectively. In the simulations, the two physics at the fundamental wavelengths and the third physics at the frequency mixing wavelength are used. In Fig. 4(d) for the non-degenerate four wave mixing, another pump beam is introduced with two fixed wavelengths at $\omega_{1}$ and $\omega_{2}$. In such situation, three physics at the fundamental wavelengths and the fourth physics at the frequency mixing wavelength are adopted. For easy comparisons, the optical intensity of input pumps are all fixed at 1 MW/cm$^{2}$ in the simulations.
\begin{figure}[htbp]
	\centering
	\includegraphics[scale=0.90]{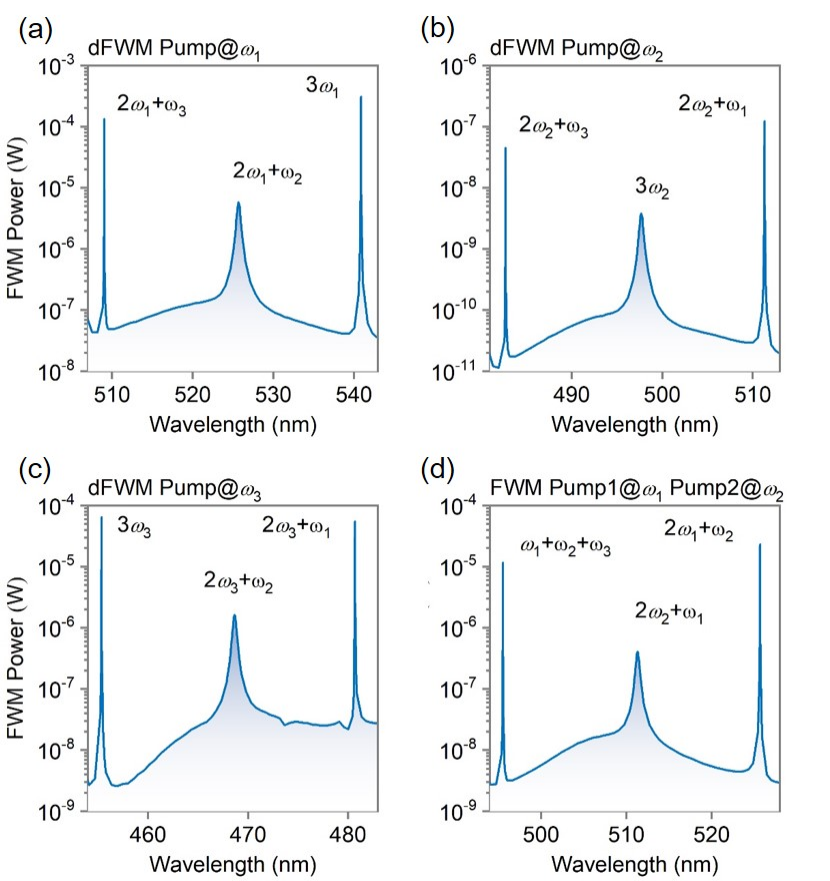}
	\caption{The nonlinearly generated visible signals originating from optical frequency mixing under $y$-polarized oblique incidence when the multiple beams simultaneously pump the metasurface. (a)-(c) Two pump beams are considered with one at a fixed wavelength of $\omega_{1}$, $\omega_{2}$, and $\omega_{3}$, respectively.  (d) Three pump beams are considered with two fixed wavelengths at $\omega_{1}$ and $\omega_{2}$.}
	\label{fig4}
\end{figure}

The generated optical mixing signals can be categorized into three groups. The first group corresponds to the THG processes when the input beams have the same frequencies. Note the THG powers here are about 2 orders of magnitude larger than those in Fig. 3, because the input power comes from the two input pumps here. The second group consists of the degenerate four-wave mixing (dFWM) processes, $2\omega_{1}+\omega_{3}$, $2\omega_{1}+\omega_{2}$, $2\omega_{2}+\omega_{3}$, $2\omega_{2}+\omega_{1}$, $2\omega_{3}+\omega_{2}$, $2\omega_{3}+\omega_{1}$, with peaks at  509 nm,  525.66 nm, 482.71 nm, 511.28 nm, 468.63 nm, 480.69 nm, respectively. The powers of these dFWM peaks do not simply scale as the product of $Q$ factors of the involved quasi-BIC modes. Instead, similarly with THG powers dependent on the internal electric energy, the dFWM radiations are determined by the combination of highly confined fields and spatial overlap of the quasi-BIC modes involved, for example in Fig. 4(b), the dFWM signals at $2\omega_{2}+\omega_{1}$ is $\sim$2.7 times larger than that at $2\omega_{2}+\omega_{3}$. The third group corresponds to the non-dFWM process with a new frequency of $\omega_{1}+\omega_{2}+\omega_{3}$ when the three different bumps simultaneously illuminate on the metasurface. Thanks to the simultaneous occurrence of the three quasi-BIC resonances at the input wavelengths, the FWM peak reaches $1.17\times10^{-5}$ W, and it is the first time to obtain such an enhanced non-dFWM in metasurface systems.

Finally, we study the power dependence of these third-order nonlinear processes. Fig. 5 shows a log-log plot of the output powers with respect to incident excitation power upon the metasurface. Here the THG process, dFWM, and non-dFWM  processes at new frequencies of $3\omega_{1}$, $2\omega_{1}+\omega_{2}$, and $\omega_{1}+\omega_{2}+\omega_{3}$ and different pump powers at $\omega_{1}$ are taken as typical examples from the three categories. As can be seen, the power of the non-dFWM process increases linearly with the excitation power at $\omega_{1}$ when the other pump power is held constant. Meanwhile, a quadratic dependence of dFWM process $2\omega_{1}+\omega_{2}$ on the $\omega_{1}$ pump power can be also observed, and this rule would also be suitable for other degenerate processes. For the THG processes, a slope value of 3 is observed as expected, implying a cube dependence of the output power on the excitation power. These power dependences verify the typical characteristics of third-order nonlinear effects\cite{Xu2019, Koshelev2019, Liu2019, Gandolfi2021, Sinev2021, Grinblat2017, Liu2018, Colom2019, Xu2022}. 
\begin{figure}[htbp]
	\centering
	\includegraphics[scale=0.25]{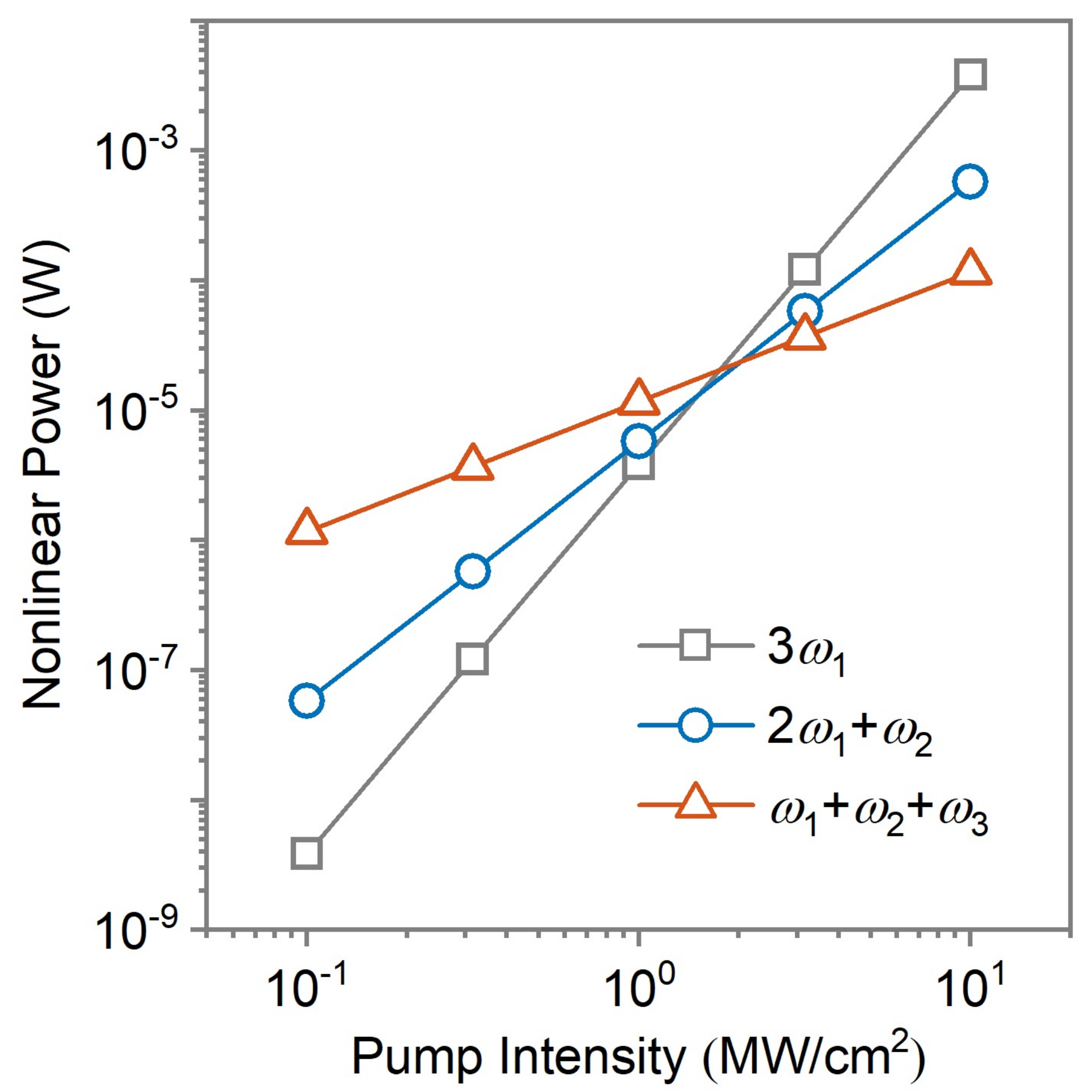}
	\caption{The power dependence of nonlinear third-order frequency mixing processes including THG ($3\omega_{1}$), dFWM ($2\omega_{1}+\omega_{2}$), and non-dFWM ($\omega_{1}+\omega_{2}+\omega_{3}$) on the power of the $\omega_{1}$ pump.}
	\label{fig5}
\end{figure}

\section{\label{sec4}Conclusions}

In summary, we have demonstrated high-efficient FWM processes in multiple-BIC resonant metasurface composed of silicon nanodisk dimers.  Under the oblique incidence, three quasi-BIC resonances with ultrahigh Q-factors are simultaneously excited in the near-infrared wavelengths, leading to the enhanced electric field and thus the large internal electric energy beneficial for boosting nonlinear effects.  Taking advantage of the multiple quasi-BIC resonances at the input wavelengths, the third-order frequency mixing processes in silicon metasurface obtain dramatic enhancement, and ten new frequencies in visible wavelengths can be observed arising from THG, dFWM and non-dFWM processes. Our work expands substantially the range of applications of BIC dielectric metasurfaces in highly efficient nonlinear metadevices, and will stimulate advances in miniaturized frequency mixing systems. Finally, it is worth pointing out that the design and mechanism in this work are general, and can be applied to other nonlinear optical effects, in particular, the photon pair generation via spontaneous parametric down-conversion\cite{Marino2019, Nikolaeva2021, Santiago-Cruz2021, Parry2021, Santiago-Cruz2022, Zhang2022}.

\begin{acknowledgments}	
	
This work is supported by the National Natural Science Foundation of China (Grants No. 11947065, No. 61901164, No. 12104105, and No. 12264028), the Natural Science Foundation of Jiangxi Province (Grant No. 20202BAB211007), the Interdisciplinary Innovation Fund of Nanchang University (Grant No. 2019-9166-27060003), the Open Project of Shandong Provincial Key Laboratory of Optics and Photonic Devices (Grant No. K202102), the Start-Up Funding of Guangdong Polytechnic Normal University (Grant No. 2021SDKYA033), and the China Scholarship Council (Grant No. 202008420045). 
	
The authors would like to thank Prof. Tingyin Ning, Prof. Lujun Huang, and Prof. Tianjing Guo for beneficial discussions on the BIC physics and nonlinear numerical simulations.
	
\end{acknowledgments}


%

\end{document}